\documentclass[aps,prl, amsmath,amssymb, twocolumn,
 groupedaddress,
 showpacs]{revtex4-1}
\usepackage{bm}
\usepackage{graphicx}
\usepackage{gensymb}
\usepackage{amsmath}
\usepackage{amssymb}
\usepackage{color}

\begin{document}

\title{Direct Observation of Nonequivalent Fermi-Arc States of Opposite Surfaces in Noncentrosymmetric Weyl Semimetal NbP}
\author{S. Souma,$^1$ Zhiwei Wang,$^2$ H. Kotaka,$^2$ T. Sato,$^3$ K. Nakayama,$^3$ Y. Tanaka,$^3$ H. Kimizuka,$^3$ \\
T. Takahashi,$^{1,3}$ K. Yamauchi,$^2$ T. Oguchi,$^2$ Kouji  Segawa,$^4$ and Yoichi Ando$^{2,5}$}
\affiliation{$^1$WPI Research Center, Advanced Institute for Materials Research, Tohoku University, Sendai 980-8577, Japan}
\affiliation{$^2$Institute of Scientific and Industrial Research, Osaka University, Ibaraki, Osaka 567-0047, Japan}
\affiliation{$^3$Department of Physics, Tohoku University, Sendai 980-8578, Japan}
\affiliation{$^4$Department of Physics, Kyoto Sangyo University, Kyoto 603-8555, Japan}
\affiliation{$^5$Institute of Physics II, University of Cologne, K\"oln 50937, Germany}

\date{\today}

\begin{abstract}
We have performed high-resolution angle-resolved photoemission spectroscopy (ARPES) on noncentrosymmetric Weyl semimetal candidate NbP, and determined the electronic states of both Nb- and P-terminated surfaces corresponding to the ``opposite" surfaces of a polar crystal. We revealed a drastic difference in the Fermi-surface topology between the opposite surfaces, whereas the Fermi arcs on both surfaces are likely terminated at the surface projection of the same bulk Weyl nodes. Comparison of the ARPES data with our first-principles band calculations suggests notable difference in electronic structure at the Nb-terminated surface between theory and experiment. The present result opens a platform for realizing exotic quantum phenomena arising from unusual surface properties of Weyl semimetals.
\end{abstract}

\maketitle

 Weyl semimetals (WSMs) manifest a novel quantum state of matter where the bulk conduction and valence bands cross at discrete points with linear dispersion in all the momentum ($\bf k$) directions in three-dimensional (3D) Brillouin zone (BZ), which can be viewed as a 3D analogue of graphene breaking time-reversal or space-inversion symmetry \cite{WanPRB11,BalentsPhys11,Nino83,HgCrSe,BalentsFilm11, Murakami07,Hirayama15,HosurRev13,BurkovRev15}. The band-crossing point is called Weyl node that is effectively described by the Weyl equation \cite{WanPRB11,BalentsPhys11,Murakami07}, and is robust against perturbations expressed in terms of Pauli matrices. The Weyl node always comes in pairs acting as a monopole (source) or anti-monopole (sink) of Berry curvature in $\bf k$ space, associated with the positive or negative chirality, respectively \cite{Nino81A,Nino81B}. The WSMs can host many exotic physical phenomena such as anomalous Hall effects, chiral anomalies \cite{Zyuzin12,Liu13}, and magnetoelectric effects \cite{SCZhang13,Chen13,Landsteiner14,Chernodub14}.

The most intriguing prediction for WSMs is the emergence of Fermi arcs on their surfaces. Unlike purely 2D metals showing {\it closed} Fermi-surface (FS) pockets, the Fermi arcs in WSMs are disjoint, {\it open} curves [see Fig. 1(a)]; they must start and end at the projections of a pair of bulk Weyl cones of opposite chiralities onto the surface BZ, independent of surface orientations and terminations. The shape of Fermi arcs in two opposite surfaces in WSMs are not equivalent with each other due to breaking of time-reversal or inversion symmetry. Such surface state is a characteristic of WSMs, leading to the predictions of intriguing phenomena like quantum interference \cite{Hosur12} and quantum oscillations \cite{Potter14}, in which the propagation of electrons across opposite surfaces through the bulk plays an essential role. Thus, a simultaneous consideration of two opposite surfaces is necessary to understand the unusual physical properties of WSMs.

Recently, density functional theory predicted that transition-metal monopnictide family TaAs, TaP, NbAs, and NbP are WSMs with twelve pairs of Weyl nodes in bulk BZ \cite{Weng15, Huang15}. These compounds crystalize in noncentrosymmetric structure [the $I4_1md$ space group; Fig. 1(b)], distinct from the WSM candidates which break time-reversal symmetry \cite{WanPRB11,HgCrSe,BalentsFilm11}. Angle-resolved photoemission spectroscopy (ARPES) of monopnictides, in particular TaAs, confirmed the existence of Fermi arcs on anion-terminated surface and bulk Weyl nodes \cite{HasanTaAsArc15,HongTaAsArc15,HongNP15,YLChenNP15,HasanNP15,HongTaP,DLFengNbP,HasanNbP}, consistent with the theoretical prediction. To firmly establish the WSM nature of monopnictides and to build a basis for the proposed exotic phenomena, it is of particular importance to experimentally establish the fermiology of both surfaces.

In this Letter, we report ARPES result on NbP in which ultrahigh carrier mobility and extremely large magnetoresistance \cite{QCNbP_NP15,MR_NbPWang} were recently reported and thus its surface transport is of great interest. We determined the electronic states of both Nb- and P-terminated surfaces. At the P-terminated surface, we observed a tadpole-shaped FS at the BZ corner, whereas such a signature is absent at the Nb-terminated counterpart, demonstrating nonequivalent electronic states between opposite surfaces. We discuss implications of our findings in terms of the bulk Weyl nodes and physical properties related to the surface Fermi arc of WSMs.

\begin{figure*}
\includegraphics[width=6.8in]{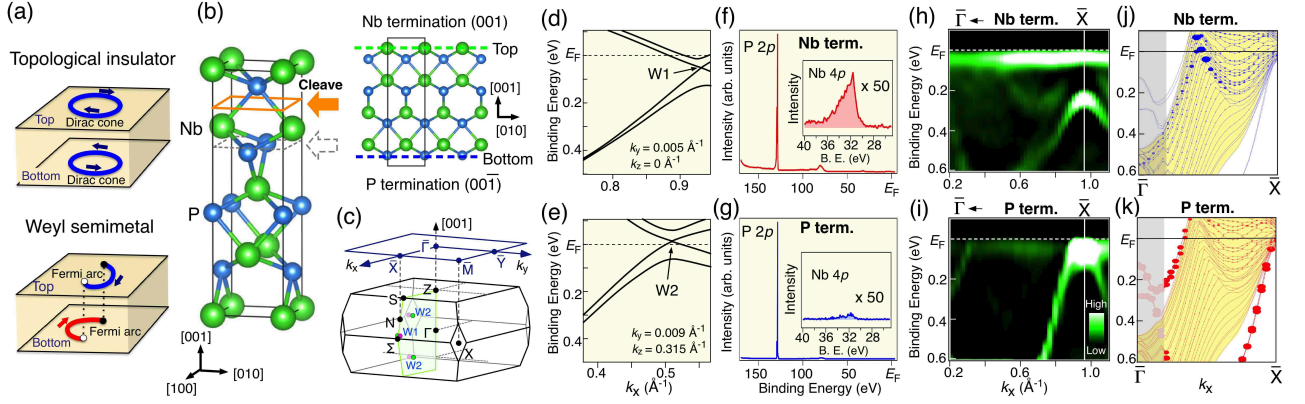}
\vspace{-0.4cm}
\caption{(Color online) (a) Schematic picture of the Dirac-cone and Fermi-arc surface states at opposite surfaces of 3D topological insulator (top) and WSM (bottom), respectively. (b) Left: crystal structure of NbP with two possible cleaving planes (arrows). Right: side view of crystal structure with Nb- and P-terminated planes at the top and bottom, respectively. (c) Bulk and surface BZs of NbP. Green shaded area is a mirror plane which is projected onto the $\bar{\Gamma}\bar{\rm X}$ line, and red and green dots highlight the Weyl nodes for W1 and W2. (d), (e) Calculated bulk-band dispersion along cuts crossing W1 and W2, respectively. (f), (g) ARPES spectrum in a wide energy region measured at  $h\nu=$ 200 eV for the Nb- and P-terminated surfaces, respectively. (h), (i) ARPES intensity along the $\bar{\Gamma}\bar{\rm X}$ cut for the Nb- and P-terminated surfaces, respectively, obtained by taking second derivative of the EDCs. (j), (k) Calculated band structure for the slab of 7-unit-cell NbP, for Nb- and P-terminated surfaces, respectively. Radius of circles represents surface spectral weight. Projection of bulk bands is shown by yellow shade. Gray shaded area in (j) and (k) is outside of the experimental band plots in (h) and (i).}
\end{figure*}

   High-quality single crystals of NbP were grown by the chemical vapor transport method. ARPES measurements were performed with the VG-Scienta SES2002 and the MBS-A1 electron analyzers with tunable synchrotron lights at the beamline BL28A at Photon Factory (KEK) and at the beamline BL7U at UVSOR. To excite photoelectrons, we used the circularly polarized lights of 50-200 eV at Photon Factory, and linearly polarized lights of 21 eV at UVSOR. The energy and angular resolution were set at 10-30 meV and 0.2$^\circ$, respectively. Samples were cleaved $in$-$situ$ along (001) crystal plane in an ultrahigh vacuum of 1$\times$10$^{-10}$ Torr. A shiny mirror-like surface was obtained after cleaving the samples, confirming its high quality.

   Electronic band-structure calculations were carried out by means of a first-principles density-functional-theory approach using projector augmented wave method implemented in Vienna $Ab$ $initio$ Simulation Package (VASP). The spin-orbit coupling was included self-consistently. For the slab calculations, we have chosen 7-unit-cell (56-layer) slab of NbP and adopted a periodic slab model with 7-layers-thick vacuum layer. To keep charge neutrality, top and bottom surfaces were terminated by the Nb and P atoms, respectively (this situation is identical to the experiment). We have calculated the weight of the wave function for the outermost 8 layers in the band-dispersion plots as well as the Fermi-surface plots to highlight the surface contribution.

We have carried out the bulk-band calculations, and confirmed that NbP is a WSM with two kinds of Weyl node pairs W1 and W2 \cite{Weng15,Huang15,YSun_calcPRB15} located near and away from the $\bar{\rm X}$ point of the surface BZ, respectively [see Fig. 1(c)]. The Weyl-cone-like dispersion can be seen from the bulk bands in Figs. 1(d) and 1(e) calculated along cuts crossing W1 or W2.
   
      At first we explain how to distinguish the Nb- and P-terminated surfaces. NbP crystal has two possible cleaving planes as indicated by broken or filled arrows in Fig. 1(b). Among these, the cleave occurs easier by breaking two Nb-P bonds per unit cell (orange line) \cite{HongNP15,YLChenNP15,YLChenNP15,YSun_calcPRB15}, rather than by breaking four bonds (gray dashed line), as supported by our calculations in which the slab with four-fold coordinated surface is 2.5 eV per unit cell more stable than that with two-fold coordination. Therefore, when we cleave the surface with [001] axis directed upward like in Fig. 1(b), the Nb-terminated (001) surface must always appear. This in return indicates that the other side of the crystal, i.e. (00$\bar{1}$) surface, is P-terminated. Such two terminations do not coexist on a single cleaved surface as long as the crystal is composed of a single domain. We accumulated the ARPES data for many cleaves, and confirmed that the obtained data are always classified into two categories attributed either to the Nb-terminated (001) or the P-terminated (00$\bar{1}$) surface, consistent with the presence of easy cleaving plane. In addition, photoelectron signals from the Nb- and P-terminated surfaces were found to never mix with each other, indicating that our single crystals are indeed of single domain.
   
\begin{figure}
\includegraphics[width=3.4in]{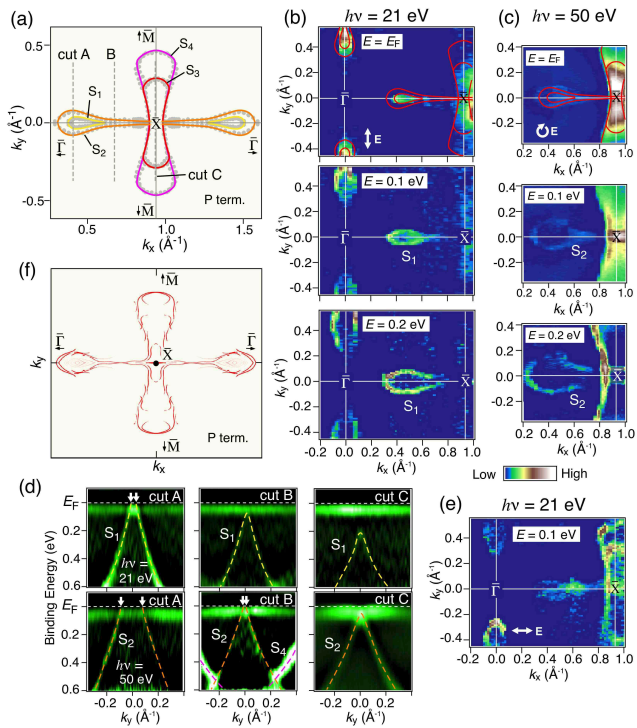}
\vspace{-0.6cm}
\caption{(Color online) 
(a) Experimental FS around the $\bar{\rm X}$ point for the P-terminated surface. Fermi wave vectors ($k_{\rm F}$'s) extracted by tracing the band dispersion at several cuts are plotted with gray dots. (b) Second-derivative intensity of the EDCs at $E_{\rm B}$ = 0.0 ($E_{\rm F}$), 0.1, and 0.2 eV measured with vertically polarized lights of $h\nu$ = 21 eV. Polarization vector of lights is indicated by arrow. Experimental FSs are overlaid for the energy slice at $E_{\rm B}$ = $E_{\rm F}$. (c) Same as (b) but measured with circularly polarized lights of $h\nu$ = 50 eV. (d) Experimental band dispersion in cuts A-C, obtained by taking second derivative of the EDCs, measured with $h\nu$ = 21 eV (top) and 50 eV (bottom). Dashed curves are a guide to the eyes to trace the band dispersions. (e) Second-derivative intensity of the EDCs at $E_{\rm B}$ = 0.1 eV measured with horizontally polarized lights of $h\nu$ = 21 eV. (f) Calculated FS for the P-terminated surface around the $\bar{\rm X}$ point. Surface weight is reflected by the gradual shading of the color scale. Momentum window is the same as (a).
}
\end{figure}
   
      Figures 1(f) and 1(g) show the energy distribution curve (EDC) of NbP for the Nb- and P-terminated surfaces, respectively, measured with $h\nu=$ 200 eV. In both surfaces, one can recognize a sharp feature at the binding energy $E_{\rm B}$ of $\sim$130 eV, which is attributed to the P 2$p$ core levels. At $E_{\rm B}$ = 32 eV, we observe another peak originating from the Nb 4$p$ core levels (inset). When we normalize the intensity of the Nb 4$p$ states with respect to that of P 2$p$, the Nb-4$p$ intensity for the P-terminated surface is much weaker than that for the Nb-terminated surface. We found that the difference in the core-level spectra is accompanied by a change in the band dispersion near $E_{\rm F}$. Figures 1(h) and 1(i) display a comparison of the ARPES intensity between two terminations along the $\bar{\Gamma}\bar{\rm X}$ cut. At the Nb-terminated surface, there exists a holelike band at $\bar{\rm X}$ which has a top of dispersion at $E_{\rm B}$ $\sim$ 0.25 eV, together with a V-shaped band within 0.3 eV of $E_{\rm F}$. On the other hand, at the P-terminated surface, the V-shaped band is absent and a holelike band touches  $E_{\rm F}$ around the $\bar{\rm X}$ point. As shown by a side-by-side comparison of experimental and calculated band dispersions in Figs. 1(i) and 1(k), the calculated surface bands reproduce overall experimental band dispersions for the P-terminated surface, whereas the experimental holelike band away from $E_{\rm F}$ is hardly reproduced for the Nb-terminated surface [Figs. 1(h) and 1(j)]. The spectral weight originating from the bulk-band projection is not well seen in the experiment, consistent with the previous ARPES studies of TaAs using vacuum-ultraviolet photons \cite{HongTaAsArc15,YLChenNP15}.

 Having established the way to experimentally distinguish the Nb- and P-terminated surfaces, a next issue is to clarify the difference in their FS topology. We begin with the electronic states of the P-terminated surface. It is noted that the anion-terminated surface has been intensively investigated by previous ARPES experiments of monopnictides, and the cross-shaped FS has been identified \cite{HasanTaAsArc15,HongTaAsArc15,HongNP15,YLChenNP15,HasanNP15,HongTaP,DLFengNbP,HasanNbP}. Figure 2(a) displays a schematic summary of the experimentally observed FS of NbP around the $\bar{\rm X}$ point. The energy bands labeled here, S$_1$-S$_4$, were found to obey different selection rules of photoelectron intensity (note that all the observed bands forming FS are the surface states \cite{HongTaAsArc15}). For example, band S$_1$ is dominantly seen at $h\nu =$ 21 eV with vertical polarization [see energy contour plots in Fig. 2(b)]. On the other hand, the intensity of outer band S$_2$ is greatly enhanced with circularly polarized 50-eV photons [Fig. 2(c)]. The intensity difference is also recognized from a comparison of the band dispersion along cut A in Fig. 2(d) where the distance between two $k_{\rm F}$ points (arrows) is obviously wider at $h\nu =$ 50 eV than that at 21 eV. Such band behavior also leads to the difference in the FS topology between S$_1$ and S$_2$; i.e., S$_1$ forms a closed pocket away from $\bar{\rm X}$ as seen from the absence of $E_{\rm F}$ crossing in cut B, whereas S$_2$ forms a tadpole FS as suggested by its $k_{\rm F}$ crossing in cut B [Fig. 2(d)]. There exist another dog-bone-shaped FSs, S$_3$ and S$_4$, elongated along the $\bar{\rm X}\bar{\rm M}$ direction as shown in Fig. 2(a), which are better resolved with horizontally polarized 21-eV photons [Fig. 2(e)]. We found that overall cross-shaped FS around the $\bar{\rm X}$ point in the experiment is reasonably reproduced by our slab calculation as shown in Fig. 2(f). In particular, $\bf k$ location of the ``head" of tadpole FS and its narrow ``tail" show good correspondence with each other (it is noted though some differences, like the absence/appearance of S$_1$ and S$_3$ pockets, can be also recognized).

\begin{figure}
\includegraphics[width=3.4in]{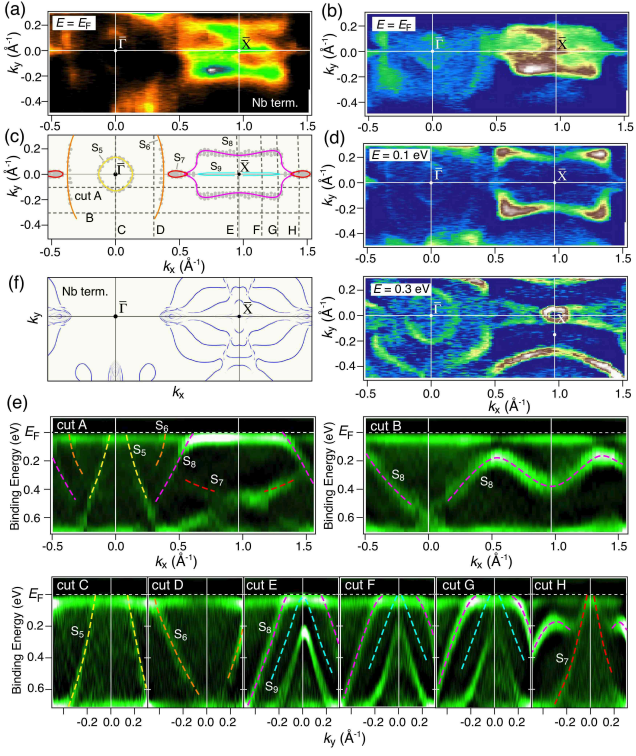}
\vspace{-0.6cm}
\caption{(Color online) (a), (b) ARPES intensity mapping at $E_{\rm F}$ and corresponding second derivative intensity of the EDCs for the Nb-terminated surface, respectively, plotted as a function of $k_x$ and $k_y$ measured with $h\nu$ = 50 eV. (c) Experimental FS for the Nb-terminated surface. $k_{\rm F}$ points extracted by tracing the band dispersion are plotted with gray dots. (d) Second-derivative intensity plots of the EDCs at $E_{\rm B}$ = 0.1 and 0.3 eV. (e) Experimental band dispersion obtained from the second derivative of the EDCs along cuts A-H in (c). Dashed curves are a guide to the eyes to trace the band dispersions. (f) Calculated FS for the Nb-terminated surface. Momentum window is the same as (c).}
\end{figure}

\begin{figure}
\includegraphics[width=3.4in]{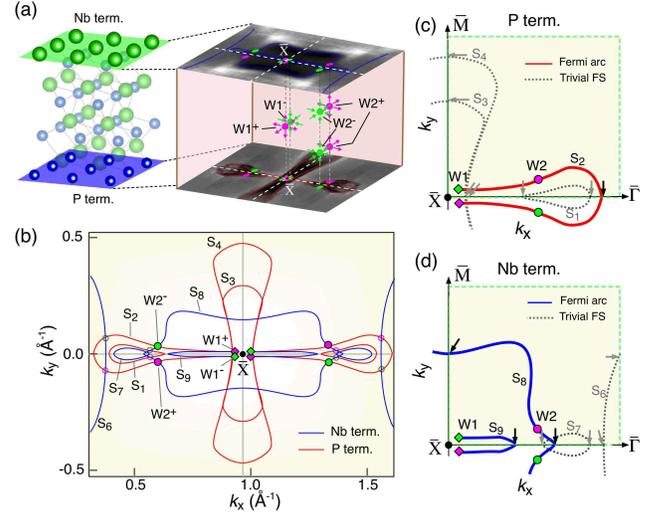}
\vspace{-1.0cm}
\caption{(Color online)
(a) Left: crystal structure of NbP highlighting the Nb- and P-terminated surfaces. Right: schematics of experimental FS around the $\bar{\rm X}$ point for the Nb- and P-terminated surfaces. Bulk Weyl nodes (W1 and W2) are illustrated by circles. Arrows indicate the Berry curvature. (b) Direct comparison of experimental FS between Nb-terminated (blue) and P-terminated (red) surfaces. Projection of Weyl nodes W2 at the intersection of FSs for opposite surfaces is overlaid by filled circles, whereas other intersections are shown by open circles. Possible Weyl pairs W1 are also indicated with diamonds. (c), (d) Schematics of Fermi arcs (solid curves) and trivial FSs (gray dashed curves) for the P- and Nb-terminated surfaces, respectively. Closed loop enclosing three Weyl nodes are indicated by green dashed rectangle. Black and gray arrows indicate the crossing points of the Fermi arcs and trivial FSs, respectively.
}
\end{figure}

   Now we turn our attention to the electronic structure of the Nb-terminated surface, which has been hardly studied in monopnictides \cite{HongTaP}. Figures 3(a) and 3(b) display ARPES-intensity mapping at $E_{\rm F}$ and corresponding second-derivative intensity of the EDCs, respectively. At first glance, FS looks much complicated and very different from that of the P-terminated surface [compare Figs. 2(a) and 3(c)]. The cross-shaped FS is absent, and the finite spectral weight is observed around $\bar{\Gamma}$. These differences are not attributable to the simple energy shift between opposite surfaces as visible from the essential difference of intensity distribution for several energy slices [compare Figs. 2(b) and 3(d)].

  The electron vs. hole-like nature of the FSs in Fig. 3(c) can be signified by tracing the experimental band dispersion at various $\bf k$ cuts shown in Fig. 3(e). Specifically, around the $\bar{\Gamma}$ point, we identify two kinds of FSs arising from bands S$_5$ and S$_6$: band S$_5$ forms a circular hole pocket, while band S$_6$ has an electronlike dispersion centered at the $\bar{\Gamma}$ point as seen from cuts A, C, and D in Fig. 3(e). Around the $\bar{\rm X}$ point, we identify a large hole pocket stemming from band S$_8$ with holelike character (cuts E-G). This large pocket is connected to a small hole pocket S$_7$ (cut H) at the midway between $\bar{\Gamma}$ and $\bar{\rm X}$. We also find a faint signature of highly anisotropic FS (S$_9$) elongated along the $\bar{\Gamma}\bar{\rm X}$ direction.
   
     The FS topology for the Nb-terminated surface is markedly different between experiment and calculation, as visible from a comparison of Figs. 3(c) and 3(f). In particular, the FS at $\bar{\Gamma}$ is absent in the calculation. Also, the shape and number of FSs around $\bar{\rm X}$ are obviously different. Possible surface reconstructions and/or enhanced interactions among surface Nb 4$d$ orbitals may need to be considered to explain such differences. We thus remark that the discrepancy between experiment and calculation should be properly taken into account in future to interpret the surface transport and spectroscopy data for the transition-metal-terminated surface of monopnictides.

Now we discuss the characteristics of observed surface states in relation to the location of bulk Weyl nodes. Figure 4(a) summarizes the present ARPES result of NbP, highlighting the significant differences between opposite surfaces. As mentioned above, there exist two kinds of bulk Weyl pairs W1$^{\pm}$ and W2$^{\pm}$ ($\pm$ represents the positive or negative chirality) in NbP \cite{Weng15,Huang15,YSun_calcPRB15}. It is expected that Weyl nodes projected onto the (001) or (00$\bar{1}$) surface terminate two Fermi arcs for W2 due to the projection from two pairs at positive and negative $k_z$'s, whereas that for W1 terminate single Fermi arc, {\it {independent of surface terminations}}. This consideration led us to experimentally pin down the $\bf k$ location of Weyl nodes, since the projection of Weyl nodes must appear at (or very close to) the intersection of FSs for opposite surfaces. While there exist three types of such intersections as indicated by filled and open circles in Fig. 4(b), filled circles are most likely due to the projection of Weyl pair W2$^{\pm}$, since (i) the calculated distance between adjacent Weyl nodes are wider for W2 than that for W1 \cite{Weng15,Huang15,YSun_calcPRB15}, and (ii) the outer tadpole FS commonly tends to form Fermi arc in monopnictides \cite{HasanTaAsArc15,HongTaAsArc15,HongNP15,YLChenNP15,HasanNP15,DLFengNbP,YSun_calcPRB15}. It turned out to be rather difficult to determine exact location of the W1 pair since the FS for opposite surfaces overlaps with each other around $\bar{\rm X}$ as visible in Fig. 4(b). Nevertheless, one can conclude that W1$^-$ and W1$^+$ are very close to each other since the FS is nearly on the $\bar{\Gamma}\bar{\rm X}$ line. In addition, the W1 pairs on both sides of $\bar{\rm X}$ are likely separated and are not traversed by the FS, since band S$_2$ appears to sink below $E_{\rm F}$ along the $\bar{\rm X}\bar{\rm M}$ cut as shown in cut C of Fig. 2(d). From these arguments, we propose in Fig. 4(c) the schematic FS for the P-terminated surface which explains most consistently the present ARPES result. The FS (S$_2$) which connects W2$^{\pm}$ can be viewed as a combination of two Fermi arcs. Namely, one arc starts and ends at W2, and the other is connected to W1. Such connectivity satisfies aforementioned rule for the number of Fermi arcs terminating the projected Weyl nodes. It is thus likely that only S$_2$ forms a Fermi arc and all the others (S$_1$, S$_3$, S$_4$) form trivial FS. At the Nb-terminated surface [Fig. 4(d)], one Fermi arc (S$_8$) starts and ends at W2, and the other is connected to another W2 pair in opposite $k_x$ region. The Weyl nodes for W1 are connected to each other via single Fermi arc S$_9$, as inferred from the absence of FS crossing along the $\bar{\rm X}\bar{\rm M}$ cut [band S$_9$ does not appear to reach $E_{\rm F}$ in cut E of Fig. 3(e)]. Thus, S$_6$ and S$_7$ are likely the trivial FSs. We emphasize, however, that the Fermi arc connectivity is not unique and depends on details of the surface \cite{Weng15}; hence, at this stage the above interpretation is just a likely possibility and its verification requires higher resolution data.

   To examine the WSM nature of NbP, we choose closed $k$-loop surrounding the odd number of (three) Weyl nodes and counted the total number of FS crossings. Since only an open Fermi arc can cross this loop an odd number of times, the total odd number of FS crossings would be a hallmark of the existence of Fermi arcs in WSM \cite{HongTaAsArc15,YLChenNP15,DLFengNbP,YSun_calcPRB15}. For the P-terminated surface [Fig. 4(c)], the Fermi arc (S$_2$) crosses this loop only once, and the trivial FSs (S$_1$, S$_3$, and S$_4$) cross six times. For the Nb-terminated surface [Fig. 4(d)], the Fermi arcs (S$_8$ and S$_9$) cross this loop three times, and the trivial FSs (S$_6$ and S$_7$) cross four times. In either case, the FSs cross the $k$-loop seven times in total, supporting the WSM nature of NbP.

 We have estimated the $k$ distance between adjacent Weyl nodes for W2 to be 0.06 $\rm{\AA}^{-1}$ in NbP, which is much smaller than that of TaAs ($\sim$ 0.15 $\rm{\AA^{-1}}$) \cite{YLChenNP15}. Such a difference is also reflected as much narrower ``tail" of tadpole FS (S$_2$) at the P-terminated surface. Since the distance between adjacent Weyl nodes is a good measure of the spin-splitting magnitude and thus the spin-orbit-coupling strength \cite{YSun_calcPRB15}, our ARPES data suggest that the spin-orbit coupling of NbP is essentially weak.
 
  The observed nonequivalent nature of the Fermi arcs between Nb- and P-terminated surfaces suggests a possibility to control the shape of Fermi arcs by tuning surface conditions, laying foundation for the Fermi-arc engineering of WSMs. It is also remarked that the nonequivalence of the surface states should be seriously taken into account in monopnictides, as long as the surface transport and spectroscopic properties, such as quantum oscillations in magnetotransport, quasiparticle interference in tunneling spectroscopy, and possible gating devices utilizing ultrathin films, are concerned.

In conclusion, we have reported ARPES results on NbP and elucidated the electronic states of the Nb- and P-terminated surfaces. We revealed that the FS topology is considerably different between these two terminations. We also found that the first-principles calculations hardly reproduce the experimental electronic structure for the Nb-terminated surface unlike the P-terminated counterpart. The present result provides a pathway for exploring new quantum phenomena utilizing Fermi-arc properties of WSMs.

\begin{acknowledgments}
We thank N. Inami, H. Kumigashira, K. Ono, S. Ideta, and K. Tanaka for their assistance in ARPES measurements. This work was supported by MEXT of Japan (Innovative Area ``Topological Materials Science"), JSPS (KAKENHI 23224010, 26287071, 25287079, 25220708, and Grant-in-Aid for JSPS Fellows 23.4376), KEK-PF (Proposal number: 2012S2-001 and 2015S2-002), and UVSOR (Proposal No. 24-536).
\end{acknowledgments}

\bibliographystyle{prsty}

\end{document}